\documentclass[sn-basic]{sn-jnl}

\usepackage{graphicx}%
\usepackage{multirow}%
\usepackage{amsmath,amssymb,amsfonts}%
\usepackage{amsthm}%
\usepackage{mathrsfs}%
\usepackage[title]{appendix}%
\usepackage{xcolor}%
\usepackage{textcomp}%
\usepackage{manyfoot}%
\usepackage{booktabs}%
\usepackage{algorithm}%
\usepackage{algorithmicx}%
\usepackage{algpseudocode}%
\usepackage{listings}%

\theoremstyle{thmstyleone}%
\theoremstyle{thmstyletwo}%

\theoremstyle{thmstylethree}%

\def\spose#1{\hbox to 0pt{#1\hss}}
\def\simlt{\mathrel{\spose{\lower 3pt\hbox{$\mathchar"218$}}
     \raise 2.0pt\hbox{$\mathchar"13C$}}}
\def\simgt{\mathrel{\spose{\lower 3pt\hbox{$\mathchar"218$}}
     \raise 2.0pt\hbox{$\mathchar"13E$}}}
\def\lsim{\rlap{$<$}{\lower 1.0ex\hbox{$\sim$}}}
\def\gsim{\rlap{$>$}{\lower 1.0ex\hbox{$\sim$}}}

\long\def\symbolfootnote[#1]#2{\begingroup%
\def\thefootnote{\fnsymbol{footnote}}\footnote[#1]{#2}\endgroup}

\begin{document}

\title{SOME MUSINGS ON ERYTHROGIGANTOACOUSTICS}

\author{Douglas Gough}
\affil{Institute of Astronomy, Madingley Road,
  Cambridge, CB3 0HA , and Department of Applied Mathematics and
  Theoretical Physics, Centre for Mathematical Sciences, Wilberforce
  Road, Cambridge, CB3 0WA, UK;\\ 
  \vskip 9pt 
  \email{douglas@ast.cam.ac.uk}
  ORCID iD:  0000-0002-6086-2636 
  DOI: xxx}

\abstract{Observations of stars other than the Sun are sensitive to oscillations of only low degree.   Many are high-order acoustic modes. 
Acoustic frequencies of main-sequence stars, for example, satisfy a well-known pattern, which some astronomers have adopted 
even for red-giant stars.   That is not wise, because the internal structures of these stars can be quite different from those on 
the Main Sequence, which is populated by stars whose structure is regular.  Here I report on pondering this matter, and point out 
two fundamental deviations from the commonly adopted relation.  There are aspects of the regular relation that are connected in 
a simple way to gross properties of the star, such as the dependence of the eigenfrequencies on the linear 
combination $n+\textstyle{\frac {1}{2}}l$ of the 
order $n$ and degree $l$, which is characteristic of a regular spherical acoustic cavity. That is not a feature of red-giant frequencies, 
because, as experienced by the waves, red-giant stars appear to have (phantom) singular centres, which 
substantially modify the 
propagation of waves.  That requires a generalization of the eigenfrequency relation, which I present here.  When fitted to 
the observed frequencies of the Sun, the outcome is consistent with the Sun being round, with no singularity in the core.  That 
is hardly novel, but at least  it provides some assurance that our understanding of stellar acoustic wave dynamics 
is on a sound footing.}


\maketitle

\section{Preamble}\label{preamble}

The first direct result of helioseismology was achieved by simple perturbation theory: estimating the correction required
of an early solar model and its acoustic frequencies, computed by  \citet{AndoOsaki1975PASJ}, to bring the frequencies  
into line with the pioneering observations of the solar $k - \omega$ diagram by \citet{deubner1975A&A}, which displays 
the relation between eigenmode 
frequencies $\omega$ and their horizontal photospheric wavenumbers $k_{\rm h}$.  Because the adjustment was relatively small, and 
because $k_{\rm h}$ was much greater than the inverse radius $R^{-1}$ of the Sun, it was expedient to adopt the precise analytical 
frequencies of a representative plane-parallel polytropic atmosphere  \citep{Lamb1911RSPSA..84..551L}, whose 
dependence on the underlying stratification is clear.  That was subsequently elucidated explicitly by \citet{SpiegelUnno1962PASJ...14...28S}, albeit for a convectively 
unstable atmosphere. The outcome of the adjustment was an increase in the mean polytropic constant in the upper boundary layer of the Sun's convection zone, which, coupled with a prior establishment of the relation between that boundary layer and the adiabat deep in the convection zone 
\citep{DOGNOW1976MNRAS} implied that the convection zone is substantially deeper than preferred estimates of the time 
\citep{DOG1977Nice}. 
Solar modellers had favoured a relatively shallow convection zone, because that is associated with a low neutrino flux, although not as low as 
the  bound inferred from the failure at the time even to detect neutrinos.  The seismic estimate was subsequently confirmed by \citet{rhodesulrichsimon1977ApJ...218..901R} with a direct numerical computation.  

That early investigation illustrates the power of analytical approximations.  In particular, it provides direct insight into the dependence 
of seismic signatures --  appropriate combinations of frequencies -- on the structure of the star, often further shedding light on 
related unresolved issues.  For example, inversion of frequency formulae derived by ray theory (either directly derived resonance conditions from propagation along the actual ray paths,
as described below, or the more flexible EBK quantization procedure \citep{Keller1958AnPhy...4..180K}, sometimes called semi-classical quantization,
even though in this case it is entirely classical).  It provides estimates of the variation of angular velocity \citep{DOG1984RSPTA.313...27G} 
 and sound speed \citep{JCDetal_1985Natur.315..378C} throughout the Sun without reference to numerically 
 computed solar models and the 
assumptions on which they rely.  Also, \citet{JCDdiagram1988IAUS} has demonstrated 
the use of the so-called large and small frequency separations, $\Delta_{n,l} = \nu_{n+1,l}-\nu_{n,l}$ and  
$\delta_{n,l} = \nu_{n,l}-\nu_{n+1,l-2}$  ($\nu_{n,l}$ being cyclic frequency of order $n$ and degree $l$),  
which respectively diagnose the large-scale structure and the 
energy-generating cores of stars, in calibrating evolved  main-sequence stellar models to provide estimates of their masses and 
central helium abundances, the latter being an indication of age.  These seismic signatures are less contaminated by extraneous 
structural properties than are the  naive wholesale  frequency comparisons 
that were carried out originally \citep[e.g.][]{JCDDOG1981A&A}.

  It is important to understand properly the role of analytical representations of oscillation frequencies in diagnosing stellar structure:  
 it is to design signatures that depend on specific structural properties with as little influence as possible of other uncertain properties 
 that the star might have.  It has been claimed that analytical formulae are necessarily inferior to (properly executed) numerical evaluations.  The latter should be more accurate, and from a sufficiently broad range of equilibrium stellar models it 
 should in principle be possible to design appropriate 
 signatures.  But in practice it has been the insight gained from the structure of analytical formulae that 
 has always provided the conduit for producing those signatures, no doubt more easily than it would have been seeking useful reliable 
 relations between raw frequencies alone, however accurate.  It must be realized that the final model calibrations would normally 
 utilize the values of those signatures determined from numerically computed eigenfrequencies, not from the approximate formulae themselves, although 
 in the early stages of an investigation it may be expedient to implement the analytical formulae directly, for that can be faster and may often  
 be more revealing.  A case in point is the first helioseismic inference which I mentioned at the outset: 
 although numerical computations of acoustic frequencies of solar models 
 had already been published \citep{RKU1970ApJ...162..993U,AndoOsaki1975PASJ}, no information was provided 
 about how those frequencies depend on pertinent properties of the solar models from 
 which they were computed.  That information was provided by the already existing valuable analytical analysis 
 by \citet{Lamb1911RSPSA..84..551L} from which the form of the $k-\omega$ diagram was evident 
 \citep[cf.][]{SpiegelUnno1962PASJ...14...28S}, and which was perfect for carrying out perturbation analysis to adequate  precision for 
 establishing that the solar convection zone was actually some 50\% deeper than what was favoured at the time.

 Also it is mandatory to interpret correctly the parameters in the analytical formulae.  For example, the oft encountered acoustic 
 travel time $\tau(R)=\int c^{-1} {\rm d}r$ from centre to seismic surface $r=R$, which determines $\omega_0$ in the asymptotic eigenfrequency relation (\ref{1.1}), has been misrepresented as being seriously ill defined  
on the grounds that 
 the integral is sensitive predominantly to the outermost layers of the star where the sound speed is low and the `surface'  appears not to be dynamically determined \citep{JNBRKU_1988RvMP...60..297B}.  
 That too is unfortunate; in this case the misunderstanding 
 arises from a failure to appreciate wave propagation in the vicinity of a singularity, such as where $c=0$.  
 In this context the seismic radius
 $R$ is a mathematical construct chosen to render its value mode-independent and simple to understand: it is essentially 
 where an appropriate 
 outward linear extrapolation of $c^2$ from the mean mode turning point vanishes, and, moreover, that extrapolation should be 
 adopted in the integrand.  It doesn't matter that in reality $c^2$ does not actually vanish; the reflected standing wave hardly senses that. 
 What it does sense is the apparent approach towards what I call a phantom singularity. 
 However, I must point out an uncertainty in its location resulting from the fact that in the vicinity of the turning point, and somewhat below, 
 the decline of $c^2$ with radius is not precisely linear as it is in a plane-parallel polytrope.  For that reason a mathematically more well defined 
 location of an acoustic seismic surface of the Sun has recently been established \citep{takataDOG2024MNRAS.527.1283T}.  Its definition is harder to  appreciate physically, and I have no need to adopt it here.
 
  Of the analytical formulae that have been adopted for helioseismology, only the low-degree versions are of use to asteroseismology.  
In particular, asymptotically high-order acoustic modes of main-sequence stars of order $n$ and degree $l$ have frequencies $\omega$ given approximately by 
\begin{equation}  \label{1.1}
\omega \simeq \left(n+\textstyle{\frac{1}{2}}l+\epsilon \right)\omega_0 - \frac{AL^2-B}{\omega}\omega_0^2
\end{equation}
\citep{tassoulasymptotics1980ApJS.43.469T,DOGEBK1986,dog1993LH}, 
where $L^2=l(l+1)$.   To this level of approximation, $\omega_0=\pi/\tau(R)$, a global property of the star, 
and the constants $\epsilon$, $A$ and $B$ also depend on only the explicit stellar structure, not on the degree nor order of the oscillation mode.  The expression for $\omega$ contains the leading terms in an expansion in 
inverse powers of 
$n$, or, equivalently, $\omega / \omega_0$; it was obtained by ignoring the Eulerian perturbations $V^\prime$ 
to the gravitational 
potential,  the  Cowling approximation \citep{Cowling1941MNRAS.101..367C}, although the neglect  of 
$V^\prime$ can have an influence on $\omega$ of similar magnitude to the second-order term   \citep{tassoulasymptoticsII1990ApJ.358.313T}.  Also ignored are rotation, acoustic glitches, Reynolds stresses 
and other small-scale fluctuations caused by convective or shear instabilities. 
The coefficient 
\begin{equation}  \label{1.2}
A=\frac{1}{2\pi \omega_0}\left(\frac{c(R)}{R}-\int_0^R \frac{1}{r}\frac{{\rm d}c}{{\rm d}r}{\rm d}r \right)
\end{equation}
is normally dominated by the sound speed in the core.  The quantities $\epsilon$ and $B$ reflect mainly the stratification in the outermost layers \citep{DOGEBK1986},  
and hardly concern me here.  This formula has been widely adopted for solar-like oscillations  \citep[e.g.][]
{JCD1984srps.conf...11C,JCDasteroseismologyreview2016arXiv160206838C,
cunhametcalfeconvcore2007ApJ...666..413C,JCDGHasteroreview2010Ap&SS.328...51C,
kallingeretal2012A&A...541A..51K}.
It has even been applied to analysing observations of Red Giants
\citep[e.g.][]{mosser_etal_2011AandA...525L...9M}, 
 a matter to which I here turn my attention.

\section{The circular drum and the homopycnic sphere}\label{drum}
Before embarking on a direct assault on the oscillation spectrum of red-giant stars, it is instructive to contemplate much simpler systems in order to garner some physical intuition about the nature of stellar acoustic modes.  I start with what is perhaps the simplest pertinent example: the uniform circular drum of radius $a$, and then a three-dimensional analogue.  

The frequencies $\omega$ of the drum are the eigenvalues of the linearized equation 
\begin{equation}  \label{2.1}
\nabla^2 \Psi + \frac{\omega^2}{c^2}\Psi = 0
\end{equation}
subject to appropriate boundary conditions, which I write as 
\begin{equation}   \label{2.2}
 \begin{split}
\Psi \;& {\rm finite} \;\; {\rm  at} \;\;  r = 0, \;\;  {\rm and} 
 \\
\frac{\partial \Psi}{\partial r}&+K \Psi = 0 \;\;\; {\rm at} \;\;\; r=a
\end{split}
\end{equation}
with respect to polar coordinates $(r,\phi)$ about the centre of the drum.
Here, $c$ is the (constant) wave phase speed.  The dependent variable $\Psi$ could be any perturbed quantity, and 
I take $K$ to be a constant.
Eigenfunctions may be represented in the separated form $\Psi(r,\phi)=\psi(r){\rm cos}(l\phi)$, in which case 
\begin{equation}   \label{2.3}
\frac{{\rm d}^2 \psi}{{\rm d}^2 r}+\frac{1}{r}\frac{{\rm d} \psi}{{\rm d} r} + \left(\frac{\omega^2}{c^2}-
\frac{l^2}{r^2}\right)\psi=0, 
\end{equation}
whose solution, regular at the origin,  is 
\begin{equation}   \label{2.4}
\psi(r)=J_l(\omega r/c),
\end{equation}
where $J_l$ is the Bessel function of the first kind of order\footnote{Yet $l$ is known as the {\it degree} of the 
eigenmode.}  $l$. 
Application of the boundary condition (\ref{2.2}) yields the eigenfrequencies, which are given asymptotically 
by 
\begin{equation}   \label{2.5}
\omega = \omega_{l,n} \sim (n+\textstyle{\frac{1}{2}}l+\epsilon)\omega_0- 
{\rm tan}^{-1}({\cal B})\pi^{-1} \omega_0 \;\;\; {\rm as} \;n \rightarrow \;\infty,
\end{equation}
where $n$ is a positive integer, $\epsilon=\frac{1}{4}$ and 
\begin{equation}   \label{2.6}
{\cal B}=\frac{8Ka -(4l^2+3)\pi}{8\pi-(4l^2-1)K\,a \omega_0^2/\omega^2}
\frac{\omega_0}{\omega} .
\end{equation}
This relation reduces to the form (\ref{1.1}) for both small and large $K$: 
\begin{equation}   \label{2.7}
\begin{split}
  {\rm if} \;K =0,  \;\; & {\rm tan}^{-1}({\cal B}) = \, \frac{(4l^2+3)\omega_0}{8\pi \omega};   
\\
  {\rm as} \;K \rightarrow \infty,  \;\; & {\rm tan}^{-1}({\cal B}) \rightarrow   \frac{(4l^2-1)\omega_0}{8\pi\omega} -\frac{1}{2}\pi .  
\end{split}
\end{equation}

It is evident that because the drum is uniform the structure of equation  (\ref{2.5}) must be determined by the 
boundary, because in this simple case the boundary is the only property of the system that incorporates the geometry.  
Its influence depends on both its shape and the phase change on reflection.  To foster appreciation 
of how the coefficients in that equation arise, it is expedient to consider how travelling waves interfere to produce 
the standing eigenfunctions.  The theory was developed by \citet{Keller1958AnPhy...4..180K} as a correction to the 
approach by \citet{Einstein1917} and \citet{Brillouin1926}  
to the Bohr-Sommerfeld quantization conditions, and is now known as EBK (sometimes semi-classical) quantization. 
It utilizes a Hamiltonian description of the waves to justify distorting ray paths into more conveniently tractable contours, and
by so doing can ease the task of recognizing the topological structure of the problem, and determining the number of  
independent quantum numbers (in this case $n$ and $l$) that fully describe the system.  Application of the procedure  
to simple classical systems, including the drum being considered here, has been presented by \citet{KellerRubinow1960AnPhy...9...24K}, and has 
subsequently been found to be a powerful tool for addressing more complicated systems, including more-star-like 
models, either spherical \citep[e.g.][]{DOGEBK1986} or not 
\citep[e.g.][]{lignieresgeorgeot.raytheory2009A&A...500.1173L}.  However, here I work with undistorted ray paths, 
because that clarifies the origins 
of some of the terms in the asymptotic frequency relation, and so provides insight into whether and how it differs 
when applied to stars with fundamentally different structures.

In the present example  of a simple drum, because $c$ is a constant, waves travel in straight lines between their reflections 
off the circular boundary, as in Figure 1(a).  
All ray segments between consecutive reflections are identical, save for a rotation,  and the angles of incidence and reflection are equal.  The rays intersect on a grazing circle, called a caustic,  surrounding a central zone of avoidance.  
At resonance the waves represented in that figure interfere to create a mode of oscillation propagating anticlockwise; a 
stationary eigenmode  is produced by the interference between two such waves propagating in opposite directions.  
As pointed out by \citet{Einstein1917}, to guarantee a regular eigenmode the raypath, namely the loci of tangents to the local wavenumber 
$\boldsymbol{k}$,  must in general 
be open, so that it samples the entire potentially reachable space; alternatively,  it could be  closed if the disc were otherwise known to be circularly symmetric, for then  an 
ensemble of space-filling rays can be found, each ray being a rotational translation of the others with appropriately  
matching phase. To achieve 
resonance, all that is necessary is that the phase transported along each ray segment matches at the surface,
or elsewhere, the 
phase of the interference pattern.  In the case of the drum, the interference pattern at the surface is simply ${\rm cos}(l\phi)$. 

A single segment of raypath, located between successive reflections, is depicted in Figure 1(b). For clarity 
I have drawn the angle $\chi$ of 
incidence at reflection to be greater than that in Figure 1(a).   Phase travels along it at the phase speed $c$, although 
there are locations, such as the boundary, where the phase can be discontinuous.  For example, if the wave function 
$\Psi$ vanishes at $r = a$, then the reflected component must simply have the opposite sign to the incident component, 
which is equivalent to a phase lag of $\pi$.  There is also a phase lag at the caustic where raypaths cross. 
That demands greater thought, which  has been provided by \citet{Keller1958AnPhy...4..180K}, who analysed 
conditions in the immediate vicinity.  The outcome 
can be pictured as follows: any ray occupies a region of width $\delta ({\rm r})$ between two adjacent  raypaths, and can conveniently be considered to take the form $\mathcal{A}(\boldsymbol{r}) {\rm exp}(\Lambda \Phi(\boldsymbol{r}))$, where 
$\Lambda \nabla \Phi = \boldsymbol{k}$ with $\mathcal{A}$ real and $\Lambda$ constant; the governing wave equation implies that for a 
stationary mode ${\rm div}(\mathcal{A}^2\nabla \Phi) = 0$, whence $\mathcal{A}^2 |\nabla \Phi |\delta = {\rm constant}$;  the crossing 
of raypaths implies that $\delta$ changes sign, which requires $\mathcal{A}^2$ to appear to change sign at a point where it is 
formally infinite;  its reality can be accommodated 
by a lag in phase $\Phi$ of $\pi /2$; that it be a lag and not an advance in phase is illustrated by, e.g.,  
\citet[][]{DOGPontedeLima2003Ap&SS.284..165G}, who describes Keller's result geometrically. 

\begin{figure*}
\centering
     \begin{center}
          \includegraphics[width=12.0cm]{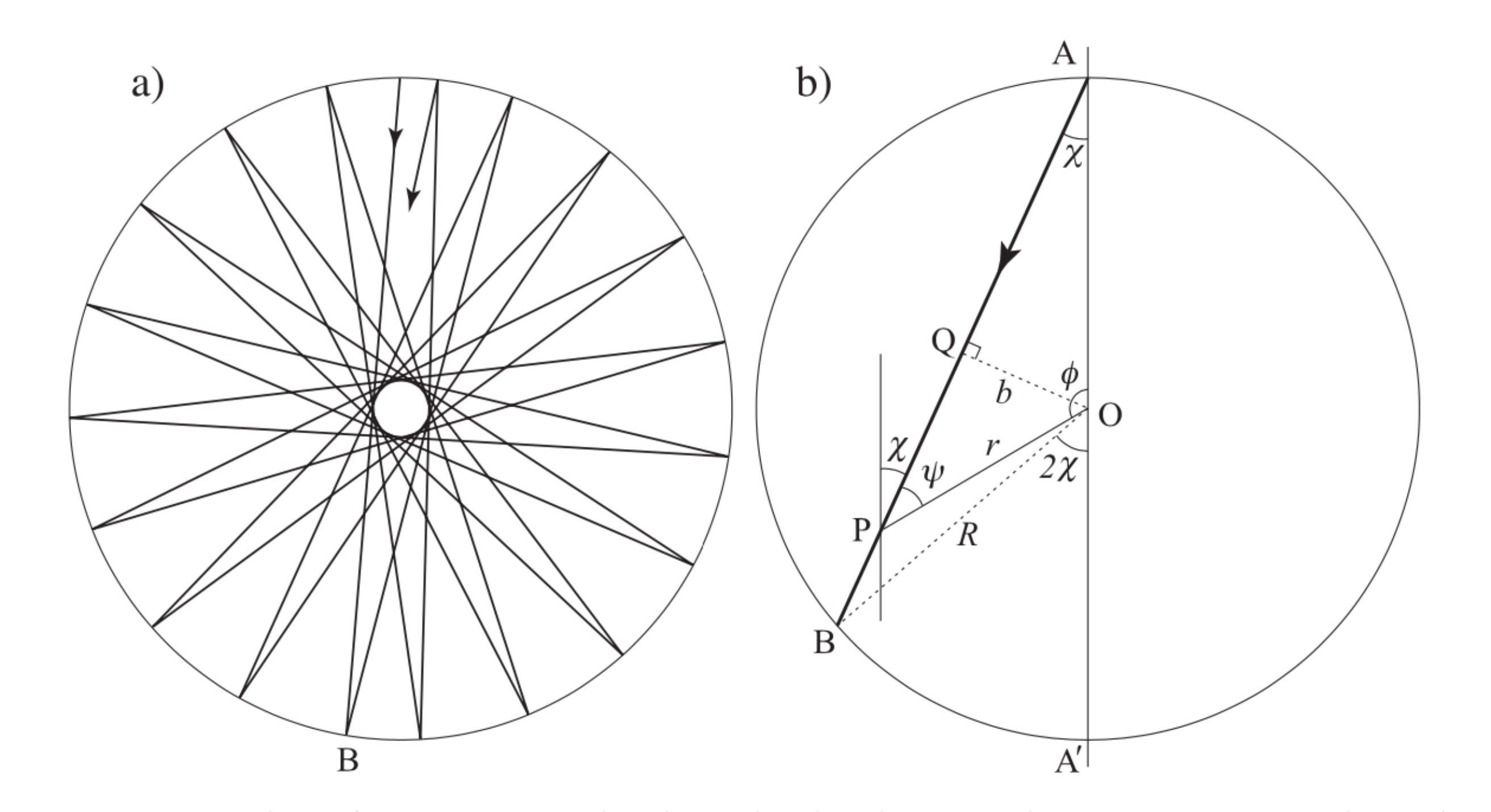}
     \end{center}
\caption{ (a) Portion of the raypath of wave propagating anticlockwise in a circular drum.  It also represents a ray in a 
plane through the centre of a homopycnic sphere.  (b) Segment of the ray illustrated in (a), but with the angle $\chi$ 
of reflection from, and incidence to, the boundary increased for clarity.  The thin fiducial line 
through P is parallel to the diameter ${\rm AOA}^\prime$. }
\label{fig1}
\end{figure*}

The resonance condition can now be established by evaluating the phase $\Phi$ of a ray at  the point B in Figure 1(b) relative to 
its fiducial value at A.  Its value is $\omega T - \textstyle \frac{3}{2}\pi$,  where $T= 2(R/c){\rm cos}\,\chi$ is the phase  
travel time between the two reflection points; subtracted from
that is the total phase lag at the boundary, namely twice half the shift at each refection, since each full shift must 
be divided equally between incident and reflected ray segments, together with the full shift resulting from the passage through the 
caustic.  Noting that at any point on the boundary there is essentially only one ray segment, the angular 
component $k_\phi$ of $\boldsymbol{k}$  must coincide with the wavenumber of the interference pattern, namely $l/R$; the 
magnitude of the wavenumber 
along the raypath AB is $k=\omega/c$.  Consequently the angle $\phi$ subtended by A and B at 
the centre of the drum O is $\phi (R) = \pi - 2 \chi$, where 
$\chi = {\rm sin}^{-1}(lc/\omega R)$, which enables $\Phi$ to be evaluated.  The phase of the interference pattern is $l\phi (R)$. 
Resonance occurs, therefore, when the two differ by an integral multiple $n'$ of $2\pi$:
\begin{equation}  \label{2.9}
\omega T - \textstyle \frac{3}{2}\pi -l\phi (R) = 2n'\pi.
\end{equation}
The fundamental  mode, namely of order $n=1$,  is characterized by having no such phase difference; therefore  $n^\prime = n-1$.
For modes of moderate degree and high order, equation (\ref{2.9}) can be expanded to yield the relation
\begin{equation}  \label{2.10}
\omega \sim \left(n+\frac{1}{2}l - \frac{1}{4}\right)\omega_0 - \frac{l^2\omega_0^2}{2\pi^2\omega} ,
\end{equation}
which is of the form of equation (\ref{1.1}). 
It is identical to the result of \citet{KellerRubinow1960AnPhy...9...24K}, who adopted the more powerful 
EBK approach.   And it is close to the exact result given by equations (\ref{2.5}) and (\ref{2.6});  indeed it is asymptotically equal to it, 
once it is recognized that having regarded the entire perturbation as an ensemble of locally sinusoidal waves the horizontally propagating 
components should also have wave numbers much greater than $R^{-1}$: for the analysis to be valid, the degree $l$ 
should formally be large compared to unity.

It is instructive to notice that the final term recorded in equation (\ref{2.10}) is dominated by the angle by which $\phi(R)$ is less than 
$\pi$.  It is offset, by half, by the corresponding diminished travel time $T$.  The phase 
jumps on reflection in the specific example considered here resulting from the requirement that the wave function vanishes, 
together with the phase retardation at the caustic circle, whose existence  is a consequence of the circular geometry of the boundary, 
determine the phase constant $\epsilon$ (in equation (\ref{1.1})).

Oscillations of a homopycnic sphere can be approached similarly.  Because reflection at the boundary is specular, raypaths lie 
in planes.  They constitute sectoral modes regarded from axes perpendicular to those planes.  The rays can be considered to 
fill tubes of cross-section $\sigma$ which, as above, satisfy  ${\rm div}(\mathcal{A}^2\nabla \Phi) = 0$, implying 
$\mathcal{A}^2 |\nabla \Phi |\sigma = {\rm constant}$, leading again to a lag in phase  $\Phi$ of $\pi /2$ at a caustic.  The phase of the 
interference pattern is related to the boundary somewhat differently: the interference pattern now occupies a disc of finite 
thickness $\delta$, determined by destructive interference between adjacent parallel raypaths encircling 
the sphere at slightly different radii 
\citep{DOGPontedeLima2003Ap&SS.284..165G}, which, from the point of view of the wave, is like a waveguide; the effective 
horizontal wavenumber is thereby increased a little, to  $(l+\textstyle{\frac{1}{2}})/r$.  A complete set of independent 
modes of degree $l$ corresponding to different azimuthal orders $m$ with respect to a given spherical polar coordinate system 
$(r,\theta,\phi)$ can be constructed as superpositions of  sectoral modes about different axes with appropriately chosen orientations. 
They all have the same frequency, of course, because the background state is spherically symmetric, and therefore 
exhibit no preferred direction.

 \section{Acoustic waves through a homobaric state}\label{homobaricstate}
 This is a somewhat artificial example, chosen just to illustrate the influence of density variation on the propagation of sound. 
 The sound speed $c$, by which I mean the speed of acoustic propagation were the density to be uniform, need not be constant. 
 For simplicity I take the classical view by considering a situation with no explicit external forcing, therefore with a homobaric 
 background state, infinite in spatial extent.  To be specific, I imagine the medium to have an equation of state satisfying 
 $\delta p = c^2 \delta \rho$  relating Lagrangian perturbations to pressure $p$ and density $\rho$.  I also restrict attention to a  
 background medium whose 
 density varies with only one Cartesian co-ordinate $x$, with scale length $H=-({\rm d\, ln}{\rho}/{\rm d}x)^{-1}$, and I consider a 
 wave of  frequency $\omega$ propagating in the $x$ direction.  
 The analysis applies approximately also to finite systems whose spatial scale of variation is much greater 
 than the wavelength of the wave\footnote{Justification for a similar extension to slowly varying background 
 states, ostensibly more familiar to astrophysicists, was explained by \citet{jeans.instability1902RSPTA.199....1J} 
 in his determination of the condition for gravitational instability in large gas clouds, and which has unjustly 
 been called `The Jeans  swindle' \citep[cf.][]{ledouxjeansinstability1951AnAp...14..438L}.}. 
 The linearized governing equations are 
 \begin{equation}  \label{3.1}
  \omega^2 \rho \xi = {{\rm d}  p'}/{{\rm d} x},
  \end{equation}
  
 \begin{equation}  \label{3.2}
  p' = \delta p = c^2 \delta\rho = -\rho c^2 {\rm d}\xi/ {\rm d}x,
\end{equation}
where $p'(x)$ is the amplitude of the Eulerian pressure perturbation and $\xi$ is displacement.  

Eliminating $\xi$ between equations (\ref{3.1}) and  (\ref{3.2}) leads to
 \begin{equation}  \label{3.3}
\frac{{\rm d}^2 p'}{{\rm d} x^2} + \frac{1}{H}\frac{{\rm d} p'}{{\rm d} x} + \frac{\omega^2}{c^2}p' =0.
\end{equation} 
This can be transformed to normal form, namely a wave equation with no first derivative, with the substitution 
$p'=\rho^{-1/2} \Psi$.  Then $\Psi$ satisfies
\begin{equation}  \label{3.5}
 \frac{{\rm d}^2 \Psi}{{\rm d} x^2} +\left( \frac{\omega^2}{c^2}-\kappa_{\rm c}^2 \right)\Psi=0 ,
 \end{equation} 
 in which $\kappa_{\rm c}$ is a critical acoustic wavenumber given by 
 \begin{equation}  \label{3.6}
 \kappa_{\rm c}^2=\frac{1}{4H^2}\left(1-2\frac{{\rm d}H}{{\rm d}x}\right) ;
 \end{equation} 
 it is usually written as $\omega_{\rm c}^2/c^2$, in terms of a critical acoustic (cut-off) frequency $\omega_{\rm c}(x)$,
 because  the latter can be compared more straightforwardly with a single global quantity, namely $\omega$, to determine where 
 in the  system the perturbation can propagate as a sinusoidal wave.   However, it is physically more revealing  
 to think in terms of  $\kappa_{\rm c}$, because only perturbations whose scale of variation    
 $\kappa_{\rm c}^{-1}$ less than the local scale length of the background state can sensibly be regarded as 
 waves which can propagate.
 
 A wave propagating obliquely to the local direction of variation of 
 $\rho$, including when that variation is not rectilinear,  satisfies
 \begin{equation} \label{3.7}
  \begin{split}
  &\nabla^2 \Psi +\left( \frac{\omega^2}{c^2}-\kappa_{\rm c}^2 \right)\Psi=0 ,
 \\
  &\kappa_{\rm c}^2 =\textstyle{\frac{1}{4}}\left(\boldsymbol{K}.\boldsymbol{K}+2\,{\rm div}\boldsymbol{K} \right) ,
 \end{split}
 \end{equation} 
 where $\boldsymbol{K}=-\rho^{-1}\nabla\rho$ is the vector inverse density scale length.   
 Unlike in the homopycnic case, the equations satisfied by the displacement  $\boldsymbol{\xi}$ are different, and 
  more complicated.  
 
 The role of $\kappa_{\rm c}$ can be appreciated physically by considering the consequences of an initial 
 perturbation to the pressure in a localized (Lagrangian) region $\cal D$ of length $\lambda$.  
 If $H$ is  substantially greater than $\lambda$, the immediate response is 
 to cause a local expansion of $\cal D$,  compressing  the adjacent environment which then reacts back on $\cal D$ 
 causing it to  contract towards its equilibrium state. 
 The compressed environment expands over a distance comparable to the size of the original 
 perturbation in a time $\tau = \lambda /c$.  The outcome mimics the original perturbation, albeit displaced.  Broadly speaking, it 
 engenders similar motion, thereby initiating a propagating wave\footnote{In this one-dimensional example 
 there are two compressed adjacent regions, one fore, the other aft, each with half the amplitude of the original 
 perturbation (if that perturbation were symmetric.)  Therefore two oppositely directed waves are produced. Why, in an initially slowly varying medium, 
 the new induced compressions do not produce substantial compressions in their wake is related to the fact that the expansion 
 of $\cal D$ overshoots the equilibrium value, thereby counteracting the trailing compression.  Cancellation is exact 
 in rectilinear motion  
 only if $H$ is infinite.  The process generalizes to two- and  three-dimensional motion.}, with cyclic frequency of order $\tau^{-1}$. 
 If, however, $H$ is less than $\lambda$, sinusoidal spatial variation hardly has meaning.  In that case, if the wave 
  is travelling in the direction of decreasing density, the material ahead has insufficient inertia to react against the 
  advancing pressure, and propagation cannot ensue.  If the density increases in the direction of propagation, 
  the wave presents too little extra pressure to compress the material ahead enough to generate a disturbance capable 
  of further propagation.    If an acoustic wave propagating in a slowly varying medium encounters a region in which 
  $\omega / c \le \kappa_{\rm c}$, the wave is therefore reflected.   Density variation plays a similar,  
  though not identical, role in the propagation of gravity waves.

  \section{Acoustically regular stars}\label{regularstars}
  By `acoustically regular' star I mean one whose dynamically pertinent state variables are essentially regular functions of $r$  (I am 
  oversimplifying and not being consistent in my terminology, as will become clear later, and I am  restricting attention 
  to spherically symmetrical systems).  Therefore the sound speed $c$ in the very core must be almost 
  uniform, for otherwise it would suffer a cusp at the centre. Elsewhere $c$ declines noticeably with $r$, and waves travelling 
  obliquely to the radial direction, interfering to form a mode of oscillation, are refracted.  This property is shared 
  by most stars, main-sequence stars in particular.  Far from the centre, waves of high order, $n \gg l$, are directed almost radially 
  through the envelope towards or away from  
  the centre of the star, so the degree of refraction is low; and it is low in the core because there the sound speed hardly varies. 
  Consequently, 
  the eigenvalue equation differs little from equation (\ref{2.5}), and indeed, as expanded upon below, retains the form of equation 
  (\ref{1.1}).  It is perhaps for this reason that equation (\ref{1.1}) has been dubbed a `universal' relation, even for red-giant stars 
  \citep[e.g.][]{mosser_etal_2011AandA...525L...9M}, although such generality has its sceptics  
  \citep[e.g.][]{HekkerJCD2017A&ARv..25....1H}.  The appellation is indeed unfortunate, as is demonstrated below.

  After separating  the wave variable $\delta p(\boldsymbol{r})$ into a function of $r$ and a spherical harmonic 
  $P^m_l ({\rm cos}\,\theta) {\rm exp}({\rm i}m\phi)$ the governing wave equation becomes
    \begin{equation} \label{4.1}
    \frac{{\rm d}^2 \Psi}{{\rm d}r^2} + \left[\frac{\omega^2}{c^2}-\kappa_{\rm c}^2-\frac{L^2}{r^2}\left(1-\frac{{\cal N}^2}{\omega^2}\right)\right]\Psi =:  \frac{{\rm d}^2 \Psi}{{\rm d}r^2} + {\cal K}^2 \Psi = 0 
    \end{equation}
     \citep[e.g.][]{dog1993LH}, in which I have recognized the role of the buoyancy frequency ${\cal N}^2$ 
     (generalized to accommodate the  spherical geometry) . Here $g$ is the local acceleration due to gravity, and now 
   \begin{equation} \label{4.2}
    \Psi = \sqrt{r^3/g \rho f}\,\delta p, 
   \end{equation} 
   where $\delta p$ now represents just the $r$-dependent amplitude of the Lagrangian pressure perturbation, and 
   where $f$ is the f-mode discriminant, given by 
    \begin{equation} \label{4.3}
    f=(\omega^2-\omega_{\rm J}^2) r/g +4-L^2g/\omega^2 r \,,
    \end{equation}
    in which $\omega_{\rm J}^2=4\pi G\rho$ is the square of the Jeans frequency.
    Propagating waves are acoustic where $f > 0$, and are gravity waves where $f<0$.  
    The buoyancy frequency, in the spherical geometry 
    of a star, is given by 
     \begin{equation} \label{4.4}
     {\cal N}^2=-g\left[{\rm d \,ln}(r^3\rho f)/{\rm d}r+g/c^2-\omega_{\rm J}^2/g \right] .
     \end{equation}
     
     Equation (\ref{4.1}) admits oscillatory solutions, representing waves, where the coefficient  ${\cal K}^2$  
     of undifferentiated 
     $\Psi$ is positive; elsewhere $\Psi$ varies more-or-less exponentially.  The radii at which that coefficient vanishes are called 
     `turning points', because they are the locations at which the wave function has a point of inflexion, turning from one kind of behaviour to the other.
     
     Before proceeding, it is important to appreciate that the critical acoustic wave number $\kappa_{\rm c}$ is significant only very close 
     to the surface, as can be recognized by noticing that if the outer layers of the envelope are approximated by a polytrope, in which $c^2$  increases approximately linearly with depth with scaleheight $H \ll R$,  
     $\kappa_{\rm c}$ is inversely proportional to depth below the phantom singularity at the acoustic surface $r=R$ where an outward linear extrapolation of $c^2$ from the upper turning point  
     vanishes, and where therefore ${\cal K}^2$ would diverge.  It causes 
     waves approaching from below to be reflected.  Therefore its affect on the 
     eigenfrequencies of deeply penetrating acoustic modes can be represented simply as a contribution to the phase factor $\epsilon$, or $\alpha$, 
     in equations (\ref{1.1}) and (\ref{4.11}), determined by conditions in a thin boundary layer beneath the acoustic surface.  
     
     Also, for high-order modes with $n \gg l$ 
     the contribution from buoyancy, through $\cal N$, can be neglected.  Therefore, beneath the surface boundary layer, waves can
      be considered to propagate at speed $c$.
     
     Several asymptotic analyses of equation (\ref{4.1}) have been presented  \citep[e.g.][]{tassoulasymptotics1980ApJS.43.469T, 
     tassoulasymptoticsII1990ApJ.358.313T, DOGEBK1986,smeyers_etal_1996A&A...307..105S}, confirming relation (\ref{1.1}).  
     From them one can 
     obtain expressions for the coefficients $\epsilon$, $A$, $B$ and $\omega_0$ for regular stars.  It is not immediately obvious 
     from them, however,  how those expressions relate physically  to the acoustical structure of the star, so here I summarize  a non-rigorous approach based directly on 
     ray tracing, which I hope will be geometrically somewhat more revealing, and which will be suggestive of whether and how equation 
     (\ref{1.1}) might need adjusting for Red Giants. 
     That it reproduces the earlier expressions adds a modicum of credulity to this more straightforward physically motivated approach.
     As with my discussion of the homopycnic sphere, I have in mind considering 
     sectoral modes, from which linear combinations can be designed to construct any other spherical harmonic.  I start from  
     Figure \ref{fig1}, which is topologically identical to its analogue pertaining here to (stratified)  regular stars.  
     The only difference is that 
     the ray paths should now be curved outwards as a result of refraction mainly near the centre 
     \citep[cf.][Figure 5]{DOGPontedeLima2003Ap&SS.284..165G}, but, as I have explained, only slightly 
     so for high-order acoustic modes.  
     
It is evident from Figure \ref{fig1} that, because the thin line through the point P on the raypath is parallel to $\rm AA^\prime$, 
\begin{equation} \label{4.5}
\phi + \psi + \chi = \pi .
\end{equation} 
Also 
\begin{equation} \label{4.6}
\begin{split}
{\rm sin}\psi=\frac{k_\phi}{k}= &\frac{Lc}{\omega r},\;\;\;{\rm and}
\\
{\rm tan}\psi=\frac{k_\phi}{k_r}= &r\frac{{\rm d}\phi}{{\rm d}r}\;\;\;{\rm along \;the\; ray} \\
=&\frac{Lc/\omega r}{\sqrt{1-L^2 c^2/\omega^2 r^2}}\,,
\end{split}
\end{equation}
where $k_r$ and $k_\phi$ are the radial and angular components of the wavenumber vector $\boldsymbol{k}$, whose magnitude is
$k=\omega/c$.
Differentiating the first of equations (\ref{4.6}) with respect to $r$ leads to  
\begin{equation} \label{4.7}
\frac{{\rm d}\psi}{{\rm d}r} = \left(\frac{{\rm d \,ln}c}{{\rm d}r}-\frac{1}{r}\right){\rm tan}\psi,
\end{equation}
which, coupled with   equation (\ref{4.5}) and the second  of equations (\ref{4.6}), yields
\begin{equation} \label{4.8}
\begin{split}
\frac{{\rm d}\chi}{{\rm d}r} =& -\frac{{\rm d \,ln}c}{{\rm d}r}{\rm tan}\psi 
\\
=& -\frac{L}{\omega}\left(1-\frac{L^2c^2}{\omega^2r^2}\right)^{-\textstyle{\frac{1}{2}}}\frac{1}{r}\frac{{\rm d \,ln}c}{{\rm d}r}\,.
\end{split}
\end{equation}
Again, from Figure \ref{fig1}, it is clear that $\psi ({\rm B}) = \chi ({\rm A}) = {\rm sin}^{-1}(Lc/ \omega R)$; $\chi ({\rm B})$ is no 
longer equal to $\chi ({\rm A})$ as it is when the raypaths are straight, as in the figure.
Integrating  equations (\ref{4.7}) and (\ref{4.8}) between the lower and upper turning points permits the establishment 
of an expression for the angle subtended between consecutive reflections at the surface $r=R$:
\begin{equation} \label{4.9} 
\phi({\rm B}) = \pi - \psi ({\rm B}) - \chi ({\rm B}) = \pi - 2\psi ({\rm B}) + \chi ({\rm A}) - \chi ({\rm B})\,.
\end{equation}
The desired resonance condition can then be written
\begin{equation} \label{4.10}
\omega T -\alpha \pi -L\phi({\rm B}) \sim 2n\pi\,,
\end{equation}
in which I have  inroduced a phase factor $\alpha$, which can be $\omega$-dependent,  
to accommodate the effect of the surface boundary layer within which 
$\kappa_{\rm c}^2$ asserts its influence.  Hence
\begin{equation} \label{4.11}
\begin{split}
&\omega T/2\pi \sim n+\textstyle\frac{1}{2}(l+\textstyle\frac{1}{2}+\alpha)-L^2D\omega_0/\pi\omega,\;\;\;{\rm as}  \;n \rightarrow \;\infty;
\\
&D  = \omega L^{-1}{\rm sin}^{-1}\left(\frac{Lc(R)}{\omega R}\right) -  
          \int^R_{Lc/\omega}{\left(1-\frac{L^2c^2}{\omega^2r^2}\right)^{-1/2} \frac{1}{r}\frac{{\rm d}c}{{\rm d}r}{\rm d}r}\,,
\end{split}
\end{equation}
a relation whose form somewhat resembles that of equation (\ref{1.1}), because  $\textstyle{\frac{1}{2}}T$ differs only slightly from 
$\tau(R)=\int{{\rm d}r/c}$, this integral being between $r=0$ and $r=R$.  Strictly, the lower limit of the integral in equation ({\ref{4.11})
should be the lower turning point $r_1$, but the simpler limit adopted here modifies the integral by only an amount of 
higher order in $\omega_0/\omega$.
The appropriate value of the phase factor $\alpha$ is somewhat uncertain, because the stratification and dynamics of 
the upper layers of stars are particularly ill understood.  I shall elaborate a little below.  
As in the case of the circular drum, the first-order term $D\omega_0/\pi\omega$ results from the displacement $\phi({\rm B})$ of point B 
from  ${\rm A}^\prime$; its influence on $\omega$ is offset partially by the consequent reduction of $T$ below $2\tau(R)$, by a factor 
which generalizes ${\rm cos}\chi$. 
The upper limit of the integral in equation (\ref{4.11}) has been replaced by $R$, because that value is independent of the wave under 
consideration and whose deviation from the upper turning point of the waves can be accommodated by the polytropic approximation 
to the near-surface layers of the equilibrium structure of the star.  I also recommend replacing $L$ by 
$\hat{L} = l+\textstyle{\frac{1}{2}}$ 
because when using ray theory, either directly, as now, or in wave-like approximations such as JWKB to equation (\ref{4.1}), it tends 
to provide a more 
accurate result, as noticed by \citet{Kramers1926ZPhy...39..828K} and \citet{kemble37} in relation to the Shr\"odinger's equation for 
the hydrogen atom.  It was justified mathematically by \citet{Langer1937PhRv...51..669L}, and from here I call it  `Langer's adjustment'.

To complete the eigenfrequency equation it is necessary to determine the value of $T$.  That cannot be achieved solely from the geometry of Figure 1 
coupled with knowledge of conditions on the boundary.  It is necessary to incorporate conditions along the 
entire raypath.  Recalling that the raypath is a trajectory against which the wavenumber $\boldsymbol{k}$ is tangent,
\begin{equation} \label{4.12}
\begin{split}
T&=\int_A^B{\frac{\hat{ \boldsymbol{k}}\cdot {\rm d}\boldsymbol{r}}{c}},\;\;\;{\rm where\;}\hat{ \boldsymbol{k}}\;{\rm is\;the\;unit\;
vector\;parallel\;to}\;\boldsymbol{k}  ,
\\
&= 2\int_{\hat{L}c/\omega}^R{ \left( 1-\frac{\hat{L}^2 c^2}{\omega^2 r^2}\right)^{-1/2} \frac{{\rm d}r}{c}}\,.
\end{split}
\end{equation} 
It is convenient also to write $\phi({\rm B})$ as an integral along the raypath:
\begin{equation} \label{4.13}
\phi({\rm B})=2\int_{\hat{L}c/\omega}^R{\frac{{\rm d}\phi}{{\rm d}r}{\rm d}r}=2\int_{\hat{L}c/\omega}^R{ \left( 1-\frac{\hat{L}^2 c^2}{\omega^2 r^2}\right)^{-1/2} \frac{{\rm d}r}{r}}\,,
\end{equation} 
because after combining it with $\omega T$ in equation (\ref{4.10}) reduction of the integral is somewhat more straightforward.
There results
\begin{equation} \label{4.14}
\omega T - \hat{L} \phi({\rm B}) = 2\omega\int_{\hat{L}c/\omega}^R{ \left( 1-\frac{\hat{L}^2 c^2}{\omega^2 r^2}\right)^{1/2} \frac{{\rm d}r}{c}}= (2n+\alpha)\pi.
\end{equation} 
This equation is essentially equivalent to an earlier result derived from EBK quantization in which the critical acoustic wave number 
$\kappa_{\rm c}$ was taken explicitly into account \citep{DOGEBK1986}.  I repeat the essence of the reduction of the integral here, 
in a rather more cavalier fashion, to serve as a prelude to a similar reduction appropriate to red-giant oscillations.

First, I recall that the sound speed in the core of a regular star does not deviate a great deal from a constant.  
Therefore, the difference $\eta$ 
between the acoustic radius variable $\tau(r)$ and its isothermal counterpart $c/r$, namely 
\begin{equation} \label{4.15}
\tau-\frac{r}{c} = \int_0^\tau{\frac{r}{c}\frac{{\rm d}c}{{\rm d}r}{\rm d}\tau } =: \eta(r)\,,
\end{equation} 
is relatively small, so a first approximation can be established by linearization.  Whence 
\begin{equation} \label{4.16}
\left( 1-\frac{\hat{L}^2 c^2}{\omega^2 r^2}\right)^{1/2}=\left( 1-\frac{\hat{L}^2}{\omega^2 \tau^2}\right)^{1/2}+
\left( 1-\frac{\hat{L}^2}{\omega^2 \tau^2}\right)^{-1/2}\frac{\hat{L}^2 \eta}{\omega^2 \tau^3} + ... \,.
\end{equation} 
In the original analysis the star was divided into the core region, where this expansion is reasonably good, matched onto the remainder
outside, which was subjected to a different approach.  Here I simply notice that in the outer region, although the relative difference 
between $r/c$ and  $\tau$ can be substantial, the direction of propagation is so close to being vertical that 
the deviation of $(1-\hat{L}^2c^2/\omega^2 r^2)$ from unity is relatively small.  So I am happy 
to retain the form of the integrals out to the upper turning point.  The first term in the expansion (\ref{4.16}) contributes
\begin{equation} \label{4.17}
2\omega\int_{\hat{L}/\omega}^{\tau(R)}\left(1-\frac{\hat{L}^2}{\omega^2\tau^2} \right)^{1/2}{{\rm d}\tau} =\frac{2\pi\omega}{\omega_0}-\hat{L}\pi+\frac{\hat{L}^2\omega_0}{\pi \omega} 
\end{equation}
to equation (\ref{4.14}).  The second term requires more analysis.  Interchanging the order of integration yields
\begin{equation} \label{4.18}
\begin{split}
I&:=\frac{\hat{L}^2}{\omega^2}\int_{\hat{L}/\omega}^{\tau(R)}\tau^{-3}\left(1-\frac{\hat{L}^2}{\omega^2 \tau^2}\right)^{-1/2}{\rm d}\tau 
\int_0^\tau\frac{r^\prime}{c^\prime}\frac{{\rm d}c^\prime}{{\rm d}r^\prime}{\rm d}\tau^\prime\,
\\
&=\left(1-\frac{\hat{L}^2}{\omega^2\tau(R)^2}\right)^{1/2}\int^{\hat{L}/\omega}_0{\frac{r}{c}\frac{{\rm d}c}{{\rm d}r}{\rm d}\tau} 
\\
&+\int^{\tau(R)}_{\hat{L}/\omega}{\left[\left(1-\frac{\hat{L}^2}{\omega^2{\tau(R)}^2}\right)^{1/2}-\left(1-\frac{\hat{L}^2}{\omega^2\tau^2}\right)^{1/2}\right]
\frac{r}{c}\frac{{\rm d}c}{{\rm d}r}{\rm d}\tau}.
\end{split}
\end{equation}
It is safe to expand the square roots about unity.  The outcome may be written
\begin{equation} \label{4.19}
\begin{split}
I&=\left(1-\frac{\hat{L}^2}{\omega^2\tau(R)^2}\right)^{1/2}\int^{\tau(R)}_0{\frac{r}{c}\frac{{\rm d}c}{{\rm d}r}{\rm d}\tau} 
\\
&-\int^{\tau(R)}_{\hat{L}/\omega}{\frac{r}{c}\frac{{\rm d}c}{{\rm d}r}{\rm d}\tau} 
+\frac{\hat{L}^2}{2\omega^2}\int^{\tau(R)}_{\hat{L}/\omega}{\frac{1}{\tau^2}\frac{r}{c}\frac{{\rm d}c}{{\rm d}r}{\rm d}\tau} .
\end{split}
\end{equation}
Expanding further, this time in the small difference $c(R)/R - 1/\tau(R)$, results in
\begin{equation} \label{4.20}
I=-\frac{\hat{L}^2}{2\omega^2}\left(\frac{c(R)}{R}-\frac{1}{\tau(R)}-\int^R_0{\frac{1}{r}\frac{{\rm d}c}{{\rm d}r}{\rm d}r}\right)\,.
\end{equation}
Combining equations (\ref{4.16}), (\ref{4.17}) and (\ref{4.20}), and substituting into equation (\ref{4.14}), yields, approximately, the desired 
eigenfrequency  equation 
\begin{equation} \label{4.21}
\frac{\omega}{\omega_0} \sim n+{\textstyle{\frac{1}{2}}}(l+{\textstyle{\frac{1}{2}}})+{\textstyle{\frac{1}{2}}}\alpha-
\frac{\hat{L}^2}{2\pi\omega}\left(\frac{c(R)}{R}-\int^R_0{\frac{1}{r}\frac{{\rm d}c}{{\rm d}r}{\rm d}r}\right)\,,
\end{equation}
which is equation ({\ref{1.1}).

Notice that again the dominant effect of wave refraction near the centre of the star is to reduce  $\phi({\rm B})$, thereby reducing the 
effective length of the portion of the interference pattern contributing to resonance and so decreasing the resonating frequency $\omega$. 

A word or two about my term `acoustically regular':  as I have already alluded, if the outer layers of the star were approximated 
by a polytrope out to the acoustic surface $r=R$ at which ${\cal K}^2$ would diverge, that location would be a singular 
point of the wave equation (\ref{4.1}).  Of the two independent solutions, one would necessarily diverge, and must be rejected, thereby 
determining the effect of the outer boundary of the star on the solution, and essentially determining $\alpha$.  In reality that need 
not be strictly accurate: above the outer turning point the amplitude of the solution declines, and the details of the stratification 
of the medium there hardly influence the structure of $\Psi$ deep inside the star.  That structure 
really depends principally on conditions below and near the upper turning point; it is approximated by the polytropic solution in that region, 
and can hardly sense whether or not $c^2$ actually vanishes.  That is why I referred to the singularity as a phantom.  An explicit 
example of the effect of removing  it by replacing the very outer layers of the approximating  polytrope with an isothermal atmosphere  
has been presented by e.g. \citet{JCDDOG1980Natur.288..544C}. The  
deviation of the near-surface stratification of the star from that of the polytrope is not of prime importance, and the upper limit 
of the integral in the small term, second-order in $l/n$, in equation (\ref{4.1}) can justifiably be 
taken to be $R$ -- provided that a suitable polytropic expression for the integrand is adopted above the turning point, the parameters 
defining the underlying polytrope being chosen to match the actual structure in the vicinity of that point.  That leads to 
$\alpha \simeq \mu + 1/2$ for effective polytropic index $\mu$.  Other expressions have been proposed; like this one, they are all 
uncertain, so I refrain from elaborating here. 
I apologize to readers familiar with acoustic wave theory for having laboured this point;  I have done so because  
failure to appreciate the role of the phantom singularity has led in the past to grave misunderstanding.  Noting that in reality 
the wave equation has no singularity at the seismic surface, I could safely confine my definition of acoustically 
regular stars to be simply those with 
regular structure throughout.  However, I prefer to extend the terminology to exclude stars with phantom wave 
singularities at their centres.  As I now demonstrate, those include Red Giants.

 \section{The Roche stellar model}\label{Rochemodel}
 Red Giants are post-main-sequence stars with dense cores, in which hydrogen has been exhausted (and perhaps some other chemical 
 elements have been produced), surrounded by extended diffuse outer envelopes  \citep[e.g.][]{faulkner2005slfh.book..149F}.  As a crude 
 first approximation to the structure one could assume, for the purpose of representing the gravitational potential,  that the entire stellar mass $M$ is contained within that core, miniscule enough to be considered a point, 
 an approximation attributed to Edouard Roche.  Were the envelope to be a homogeneous perfect gas in radiative equilibrium with Kramers opacity 
 $\kappa \propto \rho T^{-3.5}$, then the stratification would be polytropic with exponent $\mu = 3.25$.  I adopt that approximation here.
 Then, hydrostatic equilibrium requires 
 \begin{equation} \label{5.1}
 \frac{p}{p_0}=\left(\frac{r}{R}\right)^{-(\mu+1)}, \;\;\; \frac{\rho}{\rho_0}=\left(\frac{r}{R}\right)^{-\mu},
 \end{equation}
where $p_0$ and $\rho_0$ are constants satisfying $p_0/\rho_0=GM/(\mu+1)R$ in which $G$ is the gravitational constant.  The 
sound speed is given by
 \begin{equation} \label{5.2}
c^2=c_0^2\left(\frac{r}{R}\right)^{-1}, \;\;\; c_0^2=\frac{\gamma_1 p_0}{\rho_0},
\end{equation}
and the adiabatic exponent $\gamma_1$ is presumed to be constant.

The f-mode discriminant, given by equation (\ref{4.3}) with, consistent with Roche,  $\omega_{\rm J}^2$ neglected, is 
\begin{equation} \label{5.3}
f=\sigma^2x^3+4-L^2\sigma^{-2}x^{-3}, 
\end{equation}
where $ x=r/R $ and $\sigma=\omega/{\hat\omega_0},\;\hat\omega_0^2=GM/R^3$.  
It vanishes where 
\begin{equation} \label{5.4}
x=x_{\rm tr}:=\sigma^{-2/3}\left(\sqrt{L^2+4}-2\right)^{1/3},
\end{equation}
the transition radius separating an acoustic-mode (p-mode) region in the envelope above from a gravity-mode (g-mode) region in the core beneath.  Interest here is devoted to the high-order acoustic modes, which satisfy 
$\sigma^2 x^3 \gg 1$ throughout most of the acoustic propagation region.

In order to assess how the p modes and g modes interact, it is useful to record the lower p-mode and upper g-mode turning points, $x_{\rm pt}$ and  $x_{\rm gt}$, for low-degree modes,  both where ${\cal K}^2=0$.    They bracket $x_{\rm tr}$.  Their values 
depend on $\sigma$, which approximates
the order of the p modes;   when $\sigma$ is in the range 25 -- 50,  for example, $x_{\rm pt}$ decreases from about 0.13 to 0.09 
and $x_{\rm gt}$ decreases from about  0.10 to 0.07 for dipole modes.  Of crucial interest is the decay factor 
$\hat\kappa$ over the 
evanescent region between them, namely
\begin{equation} \label{5.5}
\hat\kappa = {\rm exp}\left(-\int^{x_{\rm pt}}_{x_{\rm gt}}{|{\cal K}|{\rm d}x}\right) \simeq \frac{1.1}{\sigma}\,; 
\end{equation}
it is insensitive to degree.  It is small when $\sigma$ is large, so the g-mode influence on the p-mode frequencies is typically 
not large, except perhaps near resonance, but in any case no greater than the characteristic g-mode frequency separation.  The g-mode 
frequencies are not commensurable with the p modes, so one might expect their influence on the smooth asymptotic p-mode formula, 
which, as will become apparent below, is similar to expression (\ref{4.21}), to be only weakly affected.  I shall therefore ignore the g modes in what follows, replacing the lower p-mode boundary condition by that imposed by a dynamically 
inert core.  The resulting frequencies 
approximate those of the p modes of the full model that are well apart from potentially resonating g modes.  
In any case, without prior knowledge of the g-mode frequencies I cannot take cognizance of the actual fine detail.  
However, observations 
exhibit sharp departures from the smooth p-mode trend near the g-mode resonances, and sometimes can  
 be accounted for \citep[e.g.][]{montalbanetal.2010ApJ...721L.182M}.   When there are only few g modes with frequencies in the p-mode range, that is not so straightforward \citep{cunhaetal2019MNRAS.490..909C}.  

The full expressions for the critical acoustic wave number and the buoyancy frequency are quite complicated.  My discussion here calls 
only for approximations in the p-mode region, where generally $\sigma^2 x^3 \gg 1$. Then these are given by 
\begin{equation} \label{5.6}
\begin{split}
R^2\kappa_{\rm c}^2 \simeq \,&\textstyle{\frac{1}{4}}[\mu^2+2\mu+24(\mu-2)\sigma^{-2}x^{-3}]x^{-2} \simeq 4.27x^{-2} ,
\\
{\cal N}^2/\omega^2 \simeq \, &[\mu-6-(\mu+1)\gamma_1^{-1}+12\sigma^{-2}x^{-3}]\sigma^{-2}x^{-3}\simeq 
-5.30\sigma^{-2}x^{-3}.
\end{split}
\end{equation}
For the numerical evaluations $\mu$ was taken to be 3.25 and $\gamma_1 = 5/3$.  
It should be noted, however, that near and below  $x_{\rm pt}$, $\sigma^2 x^3$  = O(1), 
and therefore $\kappa_{\rm c}^2$ = O(${\cal K}^2$); unlike in the homopycnic sphere, the influence of the critical acoustic   
wave number is significant also near the centre of the star.  Buoyancy dominates even  nearer the centre, but that is 
where g modes are being ignored.   Where $\sigma^2 x^3 \gg 1$, the effective local 
wave number in the wave equation (\ref{4.1}) is given approximately by
\begin{equation} \label{5.7}
{\cal K}^2 = [(\mu+1)/\gamma_1]\sigma^2x-[L^2+\textstyle{\frac{1}{4}}\mu(\mu+2)]/x^2.
\end{equation}
The $r^{-1/2}$ dependence of the sound speed in the Roche model is only a guide, because in reality red-giant 
envelope masses are not negligible.  An initial improvement can be made by making a correction to the gravitational 
potential from a distributed envelope mass estimated from the uncorrected Roche model.  The resulting sound 
speed satisfies 
\begin{equation} \label{5.8}
{\rm d\,ln}\,c/{\rm d\,ln}\,r = -\hat\beta(r)
\end{equation}
in which $\hat\beta < \textstyle{\frac{1}{2}}$; in fact, $\hat\beta$ decreases as $r$ decreases, by a factor of order unity.

\section{Smoothed acoustic oscillations of the Roche model}\label{Rocheoscillations}
An observation about an obvious acoustical difference between the Roche model and a regular star is not out of 
place here, for it provides reason to anticipate a fundamental variant in the eigenfrequency equation.  Because the 
sound speed, and hence the sound-speed gradient, diverge at the origin, it is no longer the case that the 
high-frequency acoustic waves of low degree pass near the centre of the star almost undeflected; indeed the 
opposite is the case.  It is straightforward 
to demonstrate that the waves are deflected by an angle $\pi-\phi (R)$ (see Figure 1) that approaches $\pi /3$ as $\omega$ approaches infinity.  
Moreover, since most of the deflection occurs only very close to the centre, it hardly affects 
the acoustic travel time.  It is therefore 
to be expected from the resonance condition (\ref{4.11}) that the principal effect on the eigenfrequency equation is to replace the combination  $n+\textstyle{\frac{1}{2}}l$ in equation (\ref{4.21}) by $n+\textstyle{\frac{1}{3}}l$.  As will immediately  become clear, the modification is actually even more severe than that, principally on account of the 
critical acoustic wavenumber.  

Equations (\ref{4.1}) and (\ref{5.6}) can be combined to yield 
\begin{equation} \label{6.1}
   \frac{{\rm d}^2 \hat\Psi}{{\rm d}z^2} + \frac{1}{z}\frac{{\rm d} \hat\Psi}{{\rm d}z}  + \frac{4}{9} \left(\frac{R^2\omega^2}{c_0^2}- 
   \frac{\Lambda^2}{z^2}\right)\hat\Psi = 0\,, 
\end{equation}
where $\hat\Psi = r^{-1/2}\Psi$, $z=(r/R)^{3/2}$ and now 
\begin{equation} \label{Lambda}
\Lambda^2=l(l+1)+\textstyle{\frac{1}{4}}(\mu+1)^2.
\end{equation}  
The solution, regular at $r=0$, is 
\begin{equation} \label{6.2}
J_\Lambda\left[\frac{2R\omega}{3c_0}\left(\frac{r}{R}\right)^{3/2}\right]\,.
\end{equation}
Setting $\delta p \propto \Psi = 0$ at the surface $r=R$ determines the eigenfrequencies:
\begin{equation} \label{6.3}
\omega = \left(n+\textstyle{\frac{1}{3}}\Lambda - \textstyle{\frac{1}{4}}\right)\omega_0 - \frac{AL^2-B}{\omega}\omega_0^2 + ... \,,
\end{equation}
a generalization of equation (\ref{1.1}) in which, as before,  
\begin{equation} \label{6.4}
\omega_0=\frac{\pi}{\tau(R)}=\pi\left(\int_0^R\frac{{\rm d}r}{c}\right)^{-1}=\frac{3\pi c_0}{2R}\,;
\end{equation}
also
\begin{equation}\label{6.5}
A=\frac{2}{9\pi^2},\;\;\;B=-\frac{(\mu+1)^2-9/4}{18\pi^2} \,.
\end{equation}

There are significant differences between this result and the relation (\ref{4.21}) pertinent to regular stars.  The most 
obvious is the factor $\textstyle{\frac{1}{3}}\Lambda$ in the leading term of the expansion.  It deviates from the corresponding 
term in relation (\ref{4.21}) in two respects: firstly, the factor $\Lambda$ in the leading term 
is a non-negligible extension to the degree $l$ of 
the mode, and results from the importance of the critical acoustic wavenumber in the innermost regions of the star, weakening 
the degree-dependence of the location of the inner caustic sphere;  secondly, the factor by which it is multiplied is smaller than 
for regular stars because of the lesser degree-dependence of the deflection resulting
from the more severe refraction caused by the greater 
sound-speed gradient near the singularity at the centre.  The first-order second term appears to reflect that in 
equation (\ref{4.21});  however, it is noteworthy that the constant $B$ depends predominantly on the stratification near the centre of the star 
\citep[cf.][]{DOGEBK1986}, rather than near the surface.  

Noting that the transition from the homopycnic sphere to regular stars has little effect on the structure of the eigenfrequency equation, 
and  given only a minor difference between the Roche model envelope, discussed here, 
and that of a red-giant star, one might expect equation 
(\ref{6.3}) to augur the structure of the corresponding relation applicable to Red Giants.  Indeed, it does.

\section{Red-giant acoustics}
One can estimate the smoothed eigenfrequencies of a Red Giant by a procedure directly analogous to that adopted above 
for regular stars.  Before embarking, it is instructive to anticipate the result in the light of past experience.   The only 
qualitative differences from a regular star are the increased sound-speed variation near the inner caustic sphere,  
and also the influence 
of the critical acoustic wavenumber $\kappa_{\rm c}$.  Rather than passing the caustic almost without deviation,  
as do the high-frequency waves in regular stars passing through a near-isothermal core, the phantom singularity in a 
Red Giant  causes the nearly radially propagating acoustic waves to experience an 
intense lateral sound-speed gradient very near the centre of the star, inducing substantial local refraction, moderated 
by $\kappa_{\rm c}$.  

Rather than modifying the ray-theory argument of \S\ref{regularstars}, for variety I work with a JWKB representation 
of the eigenfrequency equation derived from equation (\ref{4.1}), namely   
\begin{equation}\label{7.1}
I:=\int_{r_1}^R{\left(\frac{\omega^2}{c^2} -\kappa_{\rm c}^2-
\frac{\hat{L}^2}{r^2}\right)^{1/2}{\rm d}r} \sim (n+\epsilon)\pi\,,
\end{equation}
\citep{DOGJWKB2007AN....328..273G} in which $\hat{L}=l+\textstyle{\frac{1}{2}}$;  once again, Langer's adjustment has been adopted,  
and also ${\cal N}^2$ has been ignored.  The limits of integration are the lower and upper turning points, where the integrand vanishes.   
Taking my cue from the Roche model, I assume that near the inner caustic the sound speed can be well approximated by a negative 
power of $r$, namely $c \propto r^{-\hat\beta}$, where $\hat\beta$ is now constant, and that the critical acoustic wave number closely follows the inverse of $r$.  
Thus, the acoustic radius coordinate  $\tau$ is close to $\beta r/c$, where $\beta = 1/(1+\hat\beta)$.   Explicitly, I set 
\begin{equation}\label{7.2}
r/c-\beta^{-1}\tau = -\eta(r)-\hat\beta\tau =: \hat \eta(r)\,,
\end{equation}
which is the analogue of equation (\ref{4.15}), and
\begin{equation}\label{7.3}
\kappa_{\rm c}^2 = [\textstyle{\frac{1}{4}}\hat\mu(\hat\mu+2)+\hat\chi(r)]/r^2\,,
\end{equation}
in which the constant $\hat\mu$ is the value of $-{\rm d\,ln}\,\rho / {\rm d\,ln\,}r$ at, and in the vicinity of, the inner 
turning point.
By construction, $\hat \eta$ and $\hat\chi$, which measure the deviations of  $r/c$ and $\kappa_{\rm c}^2$ of 
the Red Giant from the values in the fiducial Roche model, are relatively small near $r_1$, where the direction of the ray paths deviate substantially from vertical, and probably have little effect on 
the integrand further out where they may no longer be small.

To evaluate the integral $I$ it is convenient to transform the independent variable to $\tau$, as with a regular star, 
and linearize in the deviations $\hat \eta$ and $\hat\chi$.  That includes linearizing with respect to the difference 
$\hat \eta$ between $\tau$ and its Roche counterpart
\begin{equation}\label{7.4}
\tau_{\rm R} (r) :=\int_0^r{c_0^{-1}r^{\hat\beta}{\rm d}r} \simeq \beta r/c\,,\;\;\; {\rm where}\; \beta=(1+\hat\beta)^{-1}, 
\end {equation}
which satisfies
\begin{equation}\label{7.5}
\hat \eta =\frac{r}{c}-\beta^{-1}\tau_{\rm R}=-\int_0^r{\frac{r}{c}\frac{{\rm d} c}{{\rm d}r}{\rm d}\tau}+\tau-\beta^{-1}\tau_{\rm R}\,,
 \end {equation}
in which I may safely replace $\tau_{\rm R}$ by $\tau$ to effect the linearization of the resonance integral $I$.  
The term containing $\hat{\eta}$  in the resulting integral can be simplified by interchanging the order 
of integration; it is
\begin{equation}\label{7.6}
I_{\hat\eta}:=\frac{\beta^3\Lambda^2}{\omega}\int_{\tau_1}^{\tau_2}{\left(1-\frac{\beta^2\Lambda^2}{\omega^2 {\tau^\prime}^2}\right)^{-1/2}
\tau{^\prime}^{-3}\int_0^{\tau^\prime}{\frac{r}{c}\frac{{\rm d} c}{{\rm d}r} {\rm d}\tau}}\,, 
\end{equation} 
where $\Lambda$ is still given by equation (\ref{Lambda}), except that $\mu$ should be replaced by $\hat\mu$, and 
where $\tau_1$ and $\tau_2$ locate the lower and upper tuning points.  It may be rewritten
\begin{equation}\label{7.7}
I_{\hat\eta}\simeq\beta\omega\left[\left(1-\frac{\beta^2\Lambda^2}{\omega^2\tau(R)^2}\right)^{1/2}
\int_0^{\tau(R)}{\frac{{\rm d\,ln}\, c}{{\rm d\,ln}\,r} {\rm d}\tau}- 
\int_{\beta\Lambda/\omega}^{\tau(R)}{\left(1-\frac{\beta^2\Lambda^2}{\omega^2\tau^2}\right)^{1/2}\frac{{\rm d\,ln}\, c}{{\rm d\,ln}\,r} {\rm d}\tau}\right]\,.
\end{equation} 
This expression may be approximated further by expanding the square roots and replacing $\tau^{-2}$ by $c^2/\beta^2r^2$.  After 
incorporating the resulting approximation into the resonance integral (\ref{7.1}), and evaluating the other integrals, which is straightforward, 
the eigenfrequency equation becomes 
\begin{equation}\label{7.8}
\begin{split}
\frac{\omega}{\omega_0} = n + {\textstyle{\frac{1}{2}}}\beta\Lambda + \epsilon &- \frac{A\Lambda^2-B}{\omega}\omega_0 
+ ...\,, \;\;\;\; {\rm with} \\
A=\frac{\beta^2}{2\pi^2}\left(1-\frac{\pi}{\beta\omega_0}\int_{r_1}^R{\frac{1}{r}\frac{{\rm d}c}{{\rm d}r}{\rm d}r}\right), 
\;\;\;B&=\frac{1}{2\pi \omega_0}\int_ {r_1}^R{\left(1-\frac{\Lambda^2 c^2}{\omega^2 r^2}\right)^{-1/2}\frac{c}{r^2}\hat\chi{\rm d}r}. 
\end{split}
\end{equation} 
This is the extension to Red Giants of the so-called universal relation.   As before, the factor 
$(1-\Lambda^2c^2/\omega^2r^2)^{-1/2}$ can be replaced by unity,  notwithstanding the singularity. 
It is noteworthy that, since the coefficient $A$ is in the second-order term, the quantity $\omega_0$ 
can be replaced there by its Roche value; then, when $\hat\beta = 0$, the equation for $A$ reduces 
essentially to equation (\ref{1.2}).

\section{Is the Sun a Red Giant?}
Of course not.  But is its structure acoustically red-giant-like?  That question became open to debate when  
\citet{Hawking1971MNRAS.152...75H} suggested as a putative resolution of the solar neutrino problem that 
the Sun might harbour a $10^{-16} {\rm M}_\odot$ black hole 
at its centre.  Were the black-hole mass to be much larger, that possibility might engender a structural similarity 
to that of a red-giant star, even though its envelope is not distended in quite the same way.  What would be the 
consequences?  At that time helioseismology was not available, and it was unclear whether there could be 
consequences other than the neutrino flux that could be accessible to observation.  Could the presence of a much 
more massive black hole even be possible?  \citet{Caplan_etal_solarbh2024Ap&SS.369....8C} have 
recently discussed that possibility.  They conclude that it is extremely unlikely that a primordial black hole 
with a mass sufficient to render the Sun red-giant-like could have been captured.  But, less implausibly, a 
black hole might have seeded the formation of the Sun in the first place, although it is unclear whether the 
black hole would have remained entirely in the core throughout the Sun's  main-sequence existence
\citep{DOGInsidethestars1993ASPC...40..767G}.

Helioeismology can now shed some light on the matter.  The current status of our knowledge of the stratification  
throughout the solar envelope has been thoroughly reviewed by \citet{JCDLivingReviews2021LRSP...18....2C}: 
it is very close to the best theoretical models based on standard main-sequence stellar evolution.  That 
suggests that the Sun is actually regular in its core, although the spatial resolution of the seismically inferred 
stratification near the centre is insufficient to provide certainty.  Seismology aside,  a recent study of the 
evolution of $1 {\rm M}_\odot$  stellar models by \citet{bellingeretal2023ApJ...959..113B} has concluded that 
if the Sun were 
born about a black hole whose mass has since grown to approximately $10^{-6}  {\rm M}_\odot$ it would be compatible 
with current observation.  Helioseismic inversions for sound speed and density involve the whole range of acoustic 
modes in an attempt  to sample the entire Sun;  in so doing there is the risk of diluting the power of the modes of low 
degree, which sample predominantly the central regions of the star \citep{dogagk1993ASPC...40..541G,bellingderetal2017ApJ...851...80B,JCDstellarinversion2024ApJ...961..198B}. 
It is expedient, therefore, to adopt a seismic signature based on only low-degree modes.  I do so via the 
asymptotic formula (\ref{7.8}).

I fit to the most extensive low-degree frequencies available, namely those obtained from 8640 days of  
whole-disc observations by the Birmingham Oscillations Network (BiSON) 
\citep{BiSONfreqencies2009MNRAS.396L.100B,BiSONperformance2016SoPh..291....1H}.  Since equation (\ref{7.8}) 
is applicable only to modes of high order, I have chosen to limit the frequency fitting, by least-squares weighted by 
the inverse square of the published standard errors, to only modes of order 15 and above.  As should be expected, 
the values of the parameters obtained vary somewhat with the choice of mode set adopted.   The most pertinent parameters for 
my purpose here are $\beta$,  and also $\hat\mu$, replacing $\mu$ as it appears in the definition of $\Lambda$ 
(equation (\ref{Lambda})).  The fitted values amongst the mode sets are 
\begin{equation}\label{8.1}
\beta = 1.002 \pm 0.006 \;\;\; {\rm and} \;\;\; \hat\mu = - 0.02 \pm 0.02,
\end{equation}
being the means and standard deviations amongst the mode sets.  The formal individual uncertainties in these parameters 
for each mode set, computed directly as the standard errors in the observed frequencies, are lower. 
There are also whole-disc data from the 
GOLF project on the space mission SoHO \citep{GOLFfrequencies2000A&A...355..743T}.  These are less 
extensive than the BiSON data, but where the two overlap the results are very similar.  
The conclusion is that the low-degree frequencies are proportional to $n+\textstyle{\frac{1}{2}}l$, consistent 
with the Sun being both round and regular,  as has long been suspected \citep[e.g.][]{Vandakurov1967AZh....44..786V,
DOGInsidethesun1990ASSL..159..451G}. 
  
\section{A closing remark}\label{closing}
Although the analysis presented here is very approximate,  it does demonstrate that the dramatic structural change 
experienced by a star as it becomes a Red Giant significantly modifies the frequency pattern of seismic oscillations.  
That must alert those using the pattern in trying to identify the degrees and orders of observed modes.  
Variations in the pattern have been seen in both numerical eigenfrequency computations and observations of 
real stars \citep[e.g.][]{BeddingbetaHyd2007ApJ...663.1315B,BeddingetalRG2010ApJ...713L.176B,mosser_etal_2011AandA...525L...9M,
stelloRG2014ApJ...788L..10S}.
The variations are not uncomplicated, as is indicated by even the simple-looking formula (\ref{7.8}), 
for serious changes in structure can alter, in particular, the values of $\beta,\;A\;{\rm and}\;B$ quite significantly.  
It is not clear whether those changes can be accommodated adequately by a straightforward generalization 
of the analysis presented here.  For example, taking account of the perturbation to the gravitational potential can 
make not insubstantial frequency modifictions.   What is clear, however,  is that the formula (\ref{1.1}) is no 
longer applicable.

I thank Tim Bedding, J{\o}rgen Christensen-Dalsgaard and Michael Thompson for interesting conversations, 
and Bill Chaplin for supplying the BiSON data.

\bibliography{refs}

\section{Statements and Declarations}

\noindent Funding: the authors declare that no funds, grants, or other support were received during the preparation of this manuscript.

\noindent Competing Interests: The authors have no relevant financial or non-financial interests to disclose.

\end{document}